\documentclass{optica-article}

\journal{opticajournal} 

\articletype{Research Article}

\expandafter\let\csname equation*\endcsname\relax
\expandafter\let\csname endequation*\endcsname\relax

\usepackage{amsmath}
\usepackage{graphicx}
\usepackage{dcolumn}
\usepackage{bm}
\usepackage{hyperref}
\usepackage{url}
\usepackage{braket}
\usepackage{color}

\usepackage{enumitem}

\usepackage{titlecaps}
\Addlcwords{the of and to}

\usepackage{longtable}
\usepackage{makecell}
\usepackage{etoolbox}

\newcommand{\red}[1]{{\color{black}{#1}}}
\newcommand{\blue}[1]{{\color{black}{#1}}}

\makeatletter
\newcommand{\mainmatter}{%
  \setcounter{footnote}{0}%
  \patchcmd{\@makefntext}{\fnsymbol}{\arabic}{}{}%
  \patchcmd{\@thefnmark}{\fnsymbol}{\arabic}{}{}%
  \def\@makefnmark{\textsuperscript{\arabic{footnote}}}%
}
\makeatother

\begin{document}

\title{Canonical and Poynting currents in propagation and diffraction of structured light: tutorial}

\author{Bohnishikha Ghosh,\authormark{1} Anat Daniel,\authormark{1} Bernard Gorzkowski,\authormark{1} Aleksandr Y. Bekshaev,\authormark{2} Radek Lapkiewicz,\authormark{1} and \\ Konstantin Y. Bliokh,\authormark{3,4,5}}

\address{\authormark{1}Institute of Experimental Physics, Faculty of Physics, University of Warsaw, Ludwika Pasteura 5, 02-093 Warsaw, Poland\\
\authormark{2}Physics Research Institute, Odessa I. I. Mechnikov National University, Dvorianska~2, Odessa 65082, Ukraine\\
\authormark{3}Theoretical Quantum Physics Laboratory, Cluster for Pioneering Research, RIKEN, Wako-shi, Saitama 351-0198, Japan\\
\authormark{4}Donostia International Physics Center (DIPC), Donostia-San Sebasti\'{a}n 20018, Spain\\
\authormark{5}Centre of Excellence ENSEMBLE3 Sp. z o.o., 01-919 Warsaw, Poland}



\begin{abstract*} 
The local propagation and the energy flux in structured optical fields are often associated with the Poynting vector. However, the local phase gradient (i.e., local wavevector) in monochromatic fields in free space is described by another fundamental quantity: the canonical momentum density. Distributions of the Poynting and canonical momentum densities can differ significantly from each other in structured fields. We examine the role of these quantities in the propagation and diffraction of structured optical fields, exemplified by various circularly-polarized vortex beams carrying orbital angular momentum. We describe the canonical and Poynting momentum distributions in such beams, experimentally measure the local transverse momentum density by Shack-Hartmann wavefront sensor, and investigate fine features of the diffraction of various vortex beams on a knife-edge aperture. In all cases, the measured local momentum density and local beam evolution are consistent with the canonical momentum distribution rather than the Poynting vector. Furthermore, we introduce the local angular velocity in vortex beams and determine the universal integral $\pi$ angle of azimuthal rotation in an arbitrary (yet circularly-symmetric) propagating and diffracting vortex beam. Finally, we discuss the ``supermomentum'' and ``backflow'' effects; both of these phenomena are examples of superoscillations and are related to the properties of the canonical momentum. Our results reveal the profound role of the canonical momentum in the evolution of light and demonstrate the importance of distinguishing between it and the Poynting vector in structured light.

\end{abstract*}

\section{Introduction}

Structured (i.e., inhomogeneous) wave fields are ubiquitous in nature and modern wave-based devices. Such waves are attracting ever growing attention in optics \cite{Rubinsztein2016} and other fields of wave physics \cite{Bliokh2023JO}. For a proper understanding of structured waves and their potential applications, it is important to characterize their local dynamical and propagational properties. In this context, the {\it momentum density}, or {\it energy flux density}, or {\it wave current} becomes the chief quantity of interest. Ideally, this quantity should be: (i) well defined theoretically, (ii) measurable in experiment, (iii) determine the wave-induced force (momentum transfer) in local wave-matter interactions, and (iv) determine the propagational evolution of the wave field. However, as it often happens, instead of a single well defined quantity there are several competing alternatives with different properties and specific advantages in different types of problems. 

There is a vast literature on various aspects of electromagnetic momentum density, including numerous controversies and paradoxes \cite{Brevik1979,Pfeifer2007,Barnett2010_II,Berry2009,Bekshaev2011,Bliokh2014NC}. Here we consider only one class of problems, namely, propagation of structured {\it monochromatic} light in {\it free space}. Most textbooks on optics and electromagnetism introduce the Poynting vector as the main and only quantity characterizing the momentum or energy-flux density. Typical pictures of the structure of an inhomogeneous optical field (e.g., focused \cite{Boivin1967}, diffracted \cite{Braunbek1952}, or scattered \cite{Berry2011}) show the electric-field intensity (energy-density) distribution and streamlines of the Poynting vector as if these were the fluid-like {\it currents} determining the propagational evolution of light.    

More recently, an alternative {\it canonical} (also called `orbital') momentum density was introduced for monochromatic structured optical fields \cite{Berry2009,Bekshaev2011,Bliokh2014NC}. Compared to the Poynting vector this quantity has some advantages and disadvantages. 
The main features of the Poynting momentum density are:
\begin{itemize}
\item It is universally well-defined for arbitrary (including non-monochromatic and static) electromagnetic fields in any media (anisotropic, inhomogeneous, etc.);
\item It is not directly observable in standard optical experiments;
\item It produces difficulties in determining the spin and orbital parts of the angular momentum of light \cite{Allen1999,Bekshaev2011,Bliokh2015PR,Leader2016};  
\item It does not determine the optical force on small particles \cite{Bliokh2014NC,Bliokh2015PR}. 
\end{itemize}
In turn, the main features of the canonical momentum density are: 
\begin{itemize}
\item It is well-defined only for monochromatic optical fields in free space or isotropic media \cite{Bliokh2017PRL}; 
\item It can be directly derived from the electromagnetic field-theory Lagrangian and Noether's theorem (assuming monochromaticity and Coulomb gauge) \cite{Bliokh2013NJP}; 
\item It determines the radiation-pressure force on small absorbing-dipole (Rayleigh) particles \cite{Ashkin1983,Bliokh2014NC,Bliokh2015PR,Marques2014,Leader2016} and hence is directly measurable in experiments; 
\item It allows the well-defined separation of the orbital and spin parts of the angular momentum of light \cite{Bliokh2014NC,Bliokh2015PR,Leader2016}; 
\item It is naturally connected to quantum-mechanical momentum as a `weak' (local expectation) value of the momentum operator $-i{\bm \nabla}$ \cite{Berry2009,Bliokh2013NJP_II}. 
\end{itemize}
Note that the last item is closely related to the fluid-like picture of propagation of light (or quantum-mechanical waves) within the Madelung-Bohm treatment of wave equations \cite{Madelung1927,Bohm1987}. The streamlines of the canonical momentum current correspond to the experimentally measurable {\it Bohmian trajectories} of photons \cite{Kocsis2011,Bliokh2013NJP_II}. 

Importantly, the differences between the Poynting and canonical momenta are only {\it local}, their integral values for localized fields coincide. Yet, they coincide locally for linearly polarized (i.e., spinless) fields. Still, these local differences can be crucial and cause a number of controversies, such as the circular-polarization-dependent components of the Poynting vector orthogonal to the wavevectors in the problem \cite{Fedorov1955,Imbert1972,Bliokh2014NC,Bekshaev2015,Antognozzi2016}.

The dilemma between the \red{Poynting}-like (sometimes called `kinetic') and canonical momentum densities is not unique for electromagnetic waves; it equally appears in quantum-mechanical \cite{Gordon1928,Leader2014}, acoustic \cite{Shi2019,Burns2020}, elastic \cite{Long2018,Chaplain2022,Bliokh2022PRL}, and water-wave \cite{Bliokh2022} fields. (Incidentally, the acoustic analogue of the Poynting vector for waves in solids and fluids was derived by Umov \cite{Umov} 10 years before the seminal Poynting work \cite{Poynting1884}.) For acoustic and water waves the canonical momentum is also known as `{\it pseudomomentum}' \cite{McIntyre1981,Peierls_I,Peierls_II}, and it determines the {\it Stokes drift} of the medium particles \cite{Bremer2017,Bliokh2022,Bliokh2022PRA}. Thus, in this case the canonical momentum is directly connected to the observable {\it mechanical} motion and momentum density of the medium. Therefore, thorough understanding of the Poynting vs. canonical momentum properties of optical fields is also important for other fields of wave physics. 

Despite the great progress in the description of the Poynting vs. canonical momentum dilemma in the past decade, it still causes active debates in the structured-waves community. Furthermore, many scientists outside of this community are still unfamiliar with the problems and alternatives to the Poynting-vector description of structured light. Motivated by this, here we revisit the momentum density and propagational evolution of structured light using one of the most familiar examples: vortex beams \cite{Allen1999,Bliokh2015PR,Bekshaev2011,Bekshaev_book}. It is now well established that direct optomechanical methods using small probe particles or atoms measure local optical forces proportional to the canonical momentum density (while the contribution proportional to the Poynting vector is much weaker) \cite{Bliokh2014NC, Bekshaev2015, Antognozzi2016, Angelsky2012, Yevick2016, Hayat2015}. Moreover, `quantum weak measurements' of `Bohmian trajectories' of photons also correspond to the canonical momentum density \cite{Kocsis2011, Bliokh2013NJP_II}. 
In contrast to optomechanical and \red{standard} weak-measurement methods, here we analyse more traditional optical techniques based on (A) the Shack-Hartmann wavefront sensor and (B) propagational diffraction phenomena. 

The earlier studies of vortex beams using the Shack-Hartmann wavefront sensor \cite{Leach2006,Murphy2010} and diffraction of truncated beams \cite{Arlt2003,Hamazaki2006,Cui2012,Bekshaev2014} dealt with linearly-polarized paraxial beams and hence could not sense the difference between the Poynting and canonical momentum densities. That is why all these works mentioned the Poynting vector as the measured quantity. Here we revisit these approaches for the case of circularly-polarized vortex beams and show that in fact both the Shack-Hartmann method and diffraction evolution reveal the {\it canonical} momentum density. 
Furthermore, we analyze streamlines of the canonical momentum density (i.e., Bohmian trajectories) and their relation to the diffraction phenomena: despite the claim that ``the rotation angle of the trajectory is unrelated to the Gouy phase'' \cite{Berry2008}, we show that the {\it mean} rotation angle over all the trajectories exactly corresponds to the Gouy phase in the case of the Laguerre-Gaussian (LG) beams and always produces the same $\pi$ rotation in the $z\in (-\infty, \infty)$ range for any vortex beams. 
Finally, we briefly consider the relation of the Poynting-canonical dilemma to `{\it supermomentum}' (i.e., anomalously high local wavevector) near the vortex core \cite{Berry2009,Barnett2013} and `{\it backflow}' (i.e., local retrograde propagation) \cite{Bracken1994,Berry2010}. These phenomena are currently attracting great attention \cite{Afanasev2021,Ivanov2022,Eliezer2020,Daniel2022,BB2022,Ghosh2023} and represent examples of {\it superoscillations} \cite{Berry2019,Bliokh2013NJP_II}. We argue that these phenomena are determined by the canonical momentum, while the Poynting vector cannot produce supermomentum, although can exhibit `false backflow'.  

\section{Canonical and Poynting momentum densities in vortex beams}

The time-averaged energy density, Poynting momentum density, and canonical momentum density for monochromatic paraxial optical fields in free space can be written as \cite{BornWolf,Berry2009,Bekshaev2011,Bliokh2014NC,Bliokh2015PR}:
\begin{equation}
\label{W}
W = \frac{1}{2} \!\left( |{\bf E}|^2 + |{\bf H}|^2 \right) \simeq |{\bf E}|^2\,,
\end{equation}
\begin{equation}
\label{Poynting}
{\bm \Pi} = c^{-1} \, {\rm Re}\! \left( {\bf E}^* \times {\bf H} \right)\,,
\end{equation}
\begin{equation}
\label{P}
{\bf P} = \frac{1}{2\omega} {\rm Im}\! \left[ {\bf E}^* \cdot (\bm \nabla) {\bf E} + {\bf H}^* \cdot (\bm \nabla) {\bf H}\right] \simeq \frac{1}{\omega} {\rm Im}\! \left[ {\bf E}^* \cdot (\bm \nabla) {\bf E} \right] \,.
\end{equation}
Here ${\bf E}({\bf r})$ and ${\bf H}({\bf r})$ are the complex amplitudes of the wave electric and magnetic fields, $c$ is the speed of light, $\omega$ is the frequency, we use the Gaussian units omitting inessential common factors, and also use the equivalence of the electric and magnetic field contributions in the paraxial approximation \cite{Berry2009}. The canonical momentum density represents a natural optical counterpart of the quantum-mechanical probability current ${\bf j}\propto {\rm Im} \left( \psi^* {\bm \nabla} \psi \right) = |\psi |^2 {\bm \nabla} {\rm Arg}(\psi)$ (the electric and magnetic fields play the role of the wavefunction), which can be associated with the local expectation (weak) value of the canonical momentum operator $\hat{p} = -i {\bm \nabla}$ or with the {\it local wavevector} (phase gradient) multiplied by the intensity \cite{Berry2009,Berry2013,Bliokh2013NJP_II,Bliokh2014NC,Bliokh2015PR}. The local difference between the Poynting and canonical momentum densities is determined by the Belinfante-Rosenfeld relation involving the spin density ${\bf S}$ \cite{Berry2009,Bliokh2013NJP,Bliokh2014NC,Bliokh2015PR}:
\begin{equation}
\label{Belinfante}
{\bm \Pi} = {\bf P} + \frac{1}{2} {\bm \nabla} \times {\bf S}\,,\quad
{\bf S} =  \frac{1}{2\omega} {\rm Im}\! \left( {\bf E}^* \times {\bf E} + {\bf H}^*  \times {\bf H}\right) \simeq \frac{1}{\omega} {\rm Im}\! \left( {\bf E}^* \times {\bf E} \right) \,.
\end{equation}
This relation can also be regarded as the decomposition of the Poynting momentum into orbital (canonical) and spin parts. Since the spin part is a curl of a vector field, it does not contribute to the energy transport: its divergence vanishes identically. Therefore, both the Poynting and canonical momenta satisfy the local stationary momentum conservation law, i.e., the Poynting theorem following from Maxwell's equations: ${\bm \nabla} \cdot {\bm \Pi} = {\bm \nabla} \cdot {\bf P} =0$ \cite{Berry2009,Bliokh2013NJP}.

We now consider the well-known example of the LG vortex beams with circular polarization. Assuming propagation along the $z$ axis and the zero radial index, the transverse components of their electric field can be written as \cite{Allen1999,Bekshaev2011,Bekshaev_book}:
\begin{equation}
\label{LG}
{\bf E}_\perp = {\bf e}^\sigma \Psi_\ell\,, \quad
\Psi_\ell \propto \frac{r^{|\ell|}}{w^{|\ell|+1}} \exp\!\left[-\frac{r^2}{w^2} + \frac{ikr^2}{2R} +i\ell \varphi +ikz - i (|\ell|+1) \Phi_G \right].
\end{equation}
Here ${\bf e}^\sigma = (\bar{\bf x}+i\sigma \bar{\bf y})/\sqrt{2}$ is the unit polarization vector (the overbars denote the unit vectors along the corresponding axes) corresponding to the helicity $\sigma = \pm 1$, $(r,\varphi)$ are the polar coordinates in the $(x,y)$ plane, $\ell = 0, \pm 1, \pm 2, ...$ is the integer azimuthal index (topological charge of the vortex), $w(z) = w_0\sqrt{1+z^2/z_R^2}$ is the Gaussian-envelope radius involving the waist radius $w_0$ and the Rayleigh diffraction length $z_R = k w_0^2 / 2$ ($k= \omega / c$ is the wave number), $R(z) = z (1+ z_R^2 / z^2)$ is the radius of curvature of the wavefront, and $\Phi_G (z) = \arctan (z/z_R)$ is the Gouy phase. The small longitudinal component of the field can be determined from the equation ${\bm \nabla} \cdot {\bf E} =0$ using $\partial / \partial z \simeq ik$:
\begin{equation}
\label{Ez}
E_z \simeq i k^{-1} {\bm \nabla}_\perp \cdot {\bf E}_\perp 
\simeq \frac{i}{\sqrt{2}k}e^{i\sigma \varphi}\! \left( -\frac{\sigma\ell}{r} \Psi_\ell + \frac{\partial \Psi_\ell}{\partial r}\right),
\end{equation}
where ${\bm \nabla}_\perp = (\partial /\partial x, \partial /\partial y)$. In the paraxial approximation under consideration, the magnetic field can be written assuming that the beam consists of plane waves with the same helicity $\sigma$ (i.e., is a helicity eigenstate): 
\begin{equation}
\label{H}
{\bf H} \simeq - i \sigma {\bf E} \,.
\end{equation}
%

\begin{figure}[!t]
\begin{center}
\includegraphics[width=0.8\linewidth]{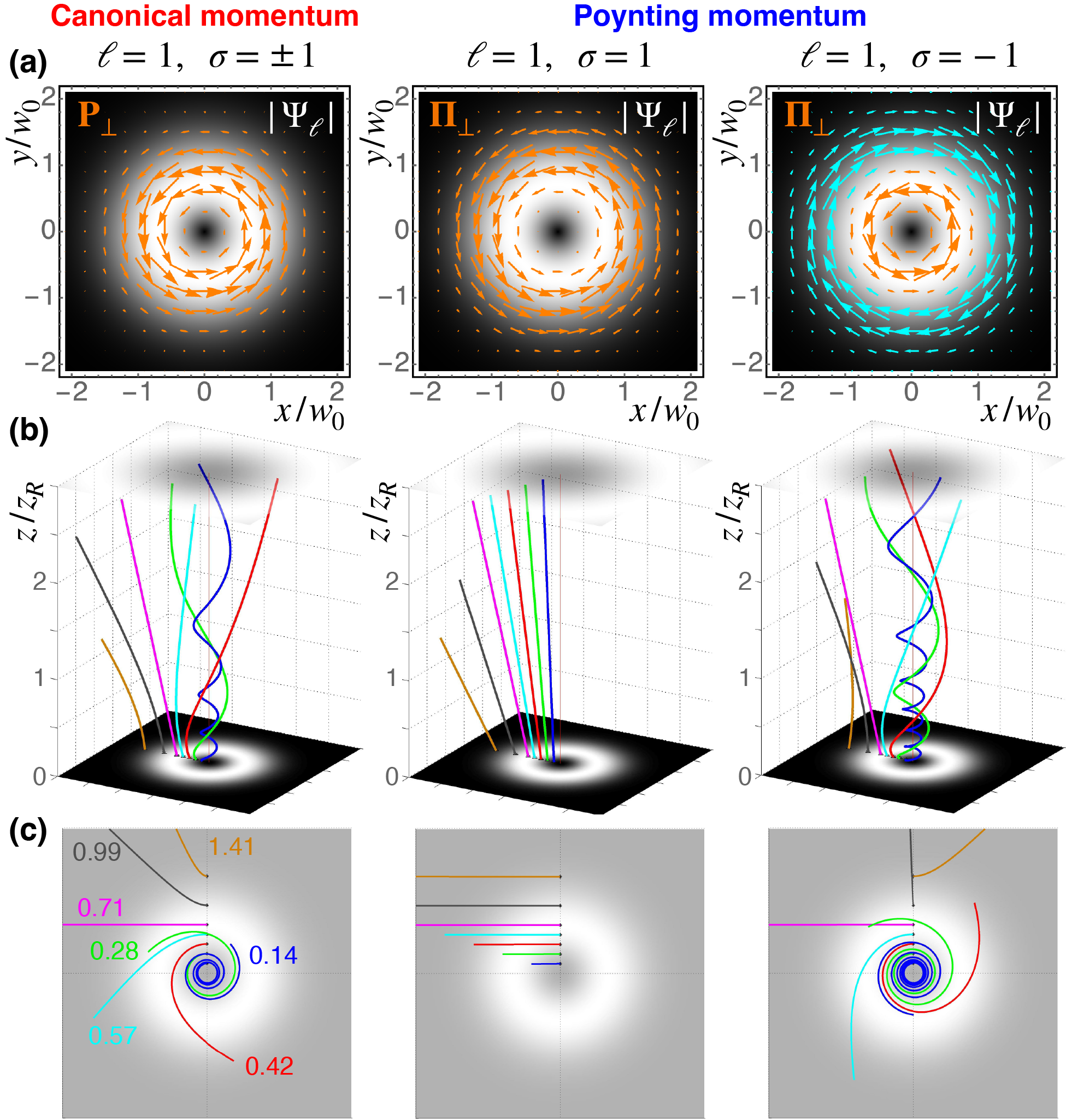}
\caption{{\bf (a)} Distributions of the transverse components of the canonical and Poynting momentum densities (\ref{P_phi}) and (\ref{Poynting_phi}), as well as of the amplitude $|\Psi_\ell |$, in the focal $z=0$ plane of the circularly-polarized LG beams (\ref{LG}) with $\ell =1$ and $\sigma=\pm 1$. The simultaneous sign flip $(\ell,\sigma) \to (-\ell,-\sigma)$ reverses the azimuthal momentum components: 
$P_\varphi \to - P_\varphi$, $\Pi_\varphi \to - \Pi_\varphi$. {\bf (b)} The streamlines $(r(z),\varphi (z))$ of the canonical and Poynting momentum densities (\ref{P_phi}) and (\ref{Poynting_phi}) in the same LG beams propagating in the $z\in (0, 3 z_R)$ range and {\bf (c)} their projections onto the transverse $(x,y)$ plane. The streamlines with the same initial azimuthal positions $\varphi_0 = \pi/2$ and different radial positions $r_0/w_0$ (indicated by coloured numbers in the lower left panel) are shown.}
\label{fig1}
\end{center}
\end{figure}

Substituting Eq.~(\ref{LG}) into Eqs.~(\ref{P}) and (\ref{W}), and neglecting quadratic contribution from the longitudinal field $E_z$, we find the canonical momentum density and energy density distributions:
\begin{equation}
\label{P_phi}
{\bf P} \simeq \frac{1}{c} \left( \frac{\ell}{k r} \bar{\bm \varphi} + \frac{r}{R} \bar{\bf r} + \bar{\bf z} \right) |\Psi_\ell |^2  \,, \quad
W \simeq |\Psi_\ell |^2\,.
\end{equation}
Here the azimuthal component of the momentum, $P_\varphi$, is responsible for the rotational dynamics in the vortex beam and the $z$-component of its angular momentum \cite{Allen1999, Bekshaev2011, Bekshaev_book, Bliokh2015PR}.
Calculation of the Poynting momentum density is more cumbersome because it involves the longitudinal field components. Substituting Eqs.~(\ref{LG})--(\ref{H}) into Eq.~(\ref{Poynting}), and using $E_r = (e^{i\sigma \varphi}/\sqrt{2}) \Psi_\ell$, $E_\varphi = i\sigma E_r$, we derive:
\begin{equation}
\label{Poynting_phi}
{\bm \Pi} \simeq \frac{1}{c} \left[ \left(\frac{\ell}{k r} - \frac{\sigma |\ell |}{k r} + \frac{2 \sigma r}{k w^2} \right)\! \bar{\bm \varphi} + \frac{r}{R} \bar{\bf r} + \bar{\bf z} \right] |\Psi_\ell |^2 \,.
\end{equation}
The Poynting momentum (\ref{Poynting_phi}) differs from the canonical momentum (\ref{P_phi}) by the $\sigma$-dependent terms in the azimuthal component. 
Note, that $\sigma$-independent and $\sigma$-dependent terms in the Poynting momentum density correspond to the canonical and spin momenta, respectively, for any circularly-polarized paraxial wave field \cite{Bekshaev2011}. In nonparaxial fields, the spin-orbit interaction effects generally produce $\sigma$-dependent terms in the canonical momentum and $\sigma$-independent terms in the spin density \cite{Bliokh2015NP, Bliokh2015PR, Bliokh2010} [see, e.g., nonparaxial Bessel beam equations (\ref{P_exact}) and (\ref{Poynting_exact}) below].

Figure~\ref{fig1}(a) shows the distributions of the transverse components of the canonical and Poynting momentum densities, as well as of the energy density, in the focal $z=0$ plane of the LG beams with $\ell = 1$ and $\sigma = \pm 1$. The $\sigma$-dependent difference between the azimuthal momentum components (\ref{P_phi}) and (\ref{Poynting_phi}) depends on the radius $r$ and generally is not small: $|\Pi_\varphi - P_\varphi | \sim |P_\varphi |$. Moreover, for $\ell\sigma < 0$ and radii $r >  w_0\sqrt{|\ell |}$, the directions of the canonical and Poynting azimuthal momentum densities are {\it opposite}: ${\rm sgn}(P_\varphi) = - {\rm sgn}(\Pi_\varphi)$. Thus, the choice of the canonical or Poynting momentum densities crucially determines the picture of the local propagational evolution of structured light. 

The streamlines of the momentum density (\ref{P_phi}) or (\ref{Poynting_phi}) \red{are the curves whose tangent vectors constitute the corresponding momentum density field. These curves can be regarded as `Bohmian trajectories' of the light propagation \cite{Padgett1995,Berry2008,Kocsis2011,Bliokh2013NJP_II}, by analogy with the streamlines of the velocity field in fluid mechanics.} Figure~\ref{fig1}(b,c) shows such streamlines $(r(z),\varphi(z))$ in the LG beams with $\ell = 1$ and $\sigma = \pm 1$ (see Appendix A for their analytical calculations). One can see that the canonical and Poynting streamlines differ dramatically from each other, except for the straight-line trajectory at the radial intensity maximum: $\partial W/ \partial r = 0$, $r_0 =  w_0 \sqrt{|\ell |/2}$. 
The difference between the canonical and Poynting momentum densities vanishes there, because, according to Eq.~(\ref{Belinfante}), it requires gradients of the spin density (proportional to the intensity for paraxial uniformly-polarized beams). 
The rectilinear propagation of the intensity-maximum points in diffracting LG beams was also noticed in other studies \cite{Bekshaev2006,Abramochkin2004}.
Note also that in the $\ell\sigma>0$ case, the interplay between the azimuthal and radial components of the Poynting vector results in globally rectilinear streamlines, see the middle panels in Fig.~\ref{fig1}(b,c).
The role of the momentum streamlines in the propagation and diffraction of vortex beams will be discussed in Section 4. 

\section{Measurements of the transverse momentum density by Shack-Hartmann sensor}

We performed experimental measurements of the radial distributions of the azimuthal component of the transverse momentum density in circularly polarized vortex beams using Shack-Hartmann wavefront sensor. \red{This is a direct purely optical method to measure the transverse components of the momentum density, without involving optomechanical effects or quantum weak measurements. It} was previously applied to linearly polarized vortex beams \cite{Leach2006}, where \red{there is no difference between the canonical and Poynting momentum densities, so that} the results were associated with the Poynting vector. Here we show that such a technique in fact measures the canonical momentum density. We used LG beams with $\ell=\pm 1$ and $\sigma=\pm 1$ in the focal plane $z=0$ where the radial component of the momentum density vanish.

\begin{figure}[!t]
\begin{center}
\includegraphics[width=0.9\linewidth]{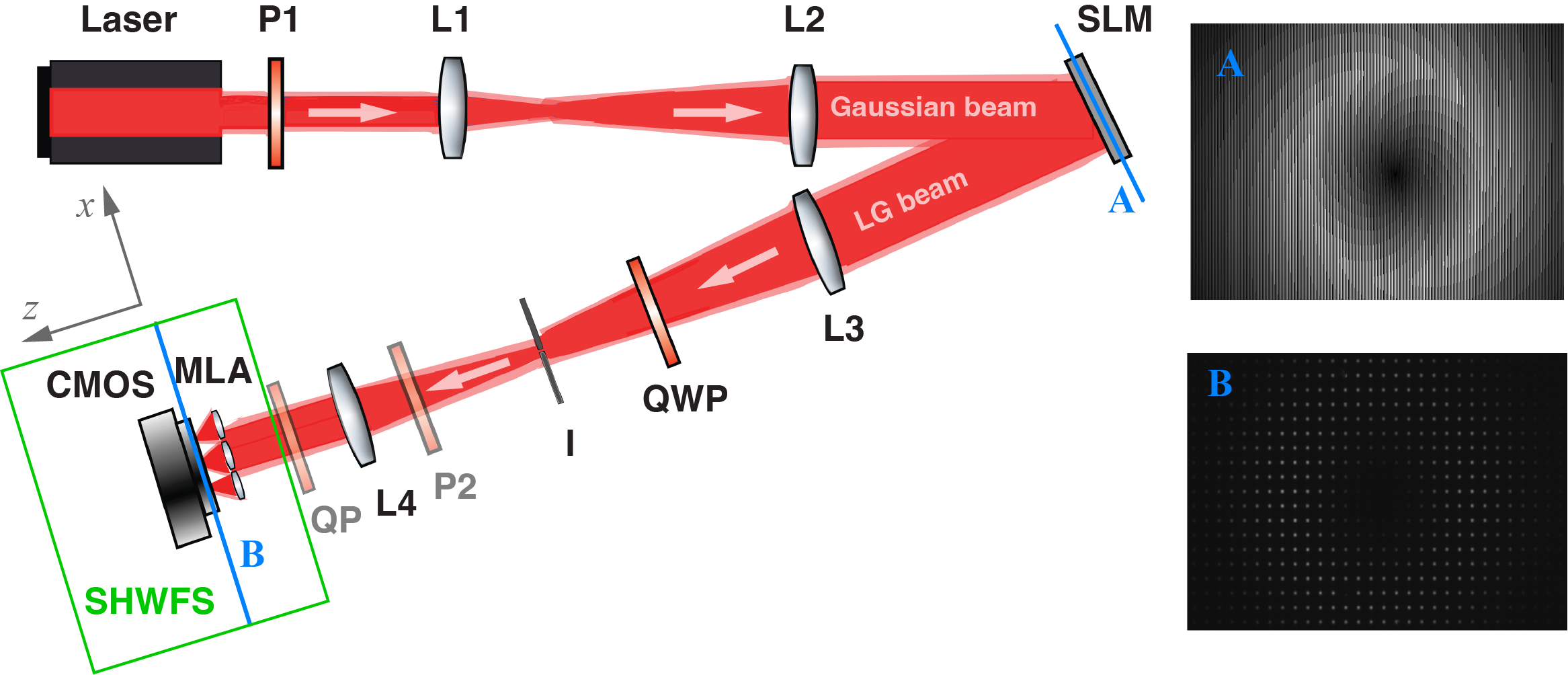}
\caption{\blue{Schematic of the experimental setup consisting of polarizers 
P1 and P2; spatial light modulator (SLM); iris (I); quarter wave plate (QWP); Q-plate (QP); micro-lens array (MLA); lenses L1 ($f= 50$~mm), L2 ($f= 500$~mm), L3 ($f= 250$~mm), and L4 ($f= 150$~mm); and complementary metal-oxide semiconductor sensor (CMOS). The Shack-Hartmann wavefront sensor (SHWFS) consists of the MLA and the CMOS. 
Inset A shows a sample hologram to produce the desired LG beam (\ref{LG}) with $|\ell|=1$. Inset B shows the corresponding spotfield observed on the CMOS sensor.}}
\label{fig2}
\end{center}
\end{figure}

The experimental setup is shown in Fig.~\ref{fig2}.
The LG beams (\ref{LG}) with $\ell = 0, \pm 1$ were produced using phase masks on a phase-only spatial light modulator (Holoeye Pluto 2.0 SLM), as shown in the inset A of Fig.~\ref{fig2}. A continuous wave laser with the wavelength $\lambda =780\,$nm (Thorlabs CLD1015) was expanded and reflected off the SLM. In order to simultaneously modulate phase and amplitude using a phase-only SLM, we adopted the technique described in \cite{Ghosh2023}, such that the desired field was obtained after filtering the first diffraction order \cite{Bolduc2013}. 
Polariser P1 was used to set the linear polarization of the incident beam.
Then, the light reflected from the SLM was circularly polarized using a quarter wave plate (QWP) oriented at $45^\circ$ with respect to P1. Polarizer P2 was used to determine the orientations of the fast and slow axes of the QWP, and it was removed after that. Also, to determine the circular-polarization helicity $\sigma$, a Q-plate (QP) of the order $q=1/2$ (Thorlabs-WPV10L-780) was employed \cite{Rubano2019} and removed afterwards. This process allowed us to create LG beams with desired circular polarization.  

Next, the SLM was imaged using lenses L3 and L4 onto the microlens array (ThorLabs-MLA-150-5C-M, each lens has a pitch of $150\,{\rm \mu m}$ and a focal length of 5.6~mm) that focused the beam onto the CMOS camera (mvBlueFOX-200wG, pixel size $6\,{\rm \mu m}$). The inset B in Fig.~\ref{fig2} shows the {\it spotfield} generated on the CMOS when the mask shown in the inset A was displayed on the SLM. 
\blue{(The spotfield is a 2D-grid-like arrangement of focal spots generated by a Shack-Hartmann wavefront sensor. This pattern is formed as each lens of the microlens array focuses the corresponding small area of the incoming wavefront onto the corresponding region on the CMOS sensor.)}
In accordance with the principle of the Shack-Hartmann sensor \cite{Kong2017}, a reference spotfield is generated by creating the hologram corresponding to a wide Gaussian beam on the SLM and subsequently imaging the light reflected from the SLM plane onto the microlens array. While generating the spotfield of the reference (a wide Gaussian beam), the polarization of the light incident on and reflected from the SLM were once again set according to the aforementioned procedure. 

\begin{figure}[!t]
\begin{center}
\includegraphics[width=\linewidth]{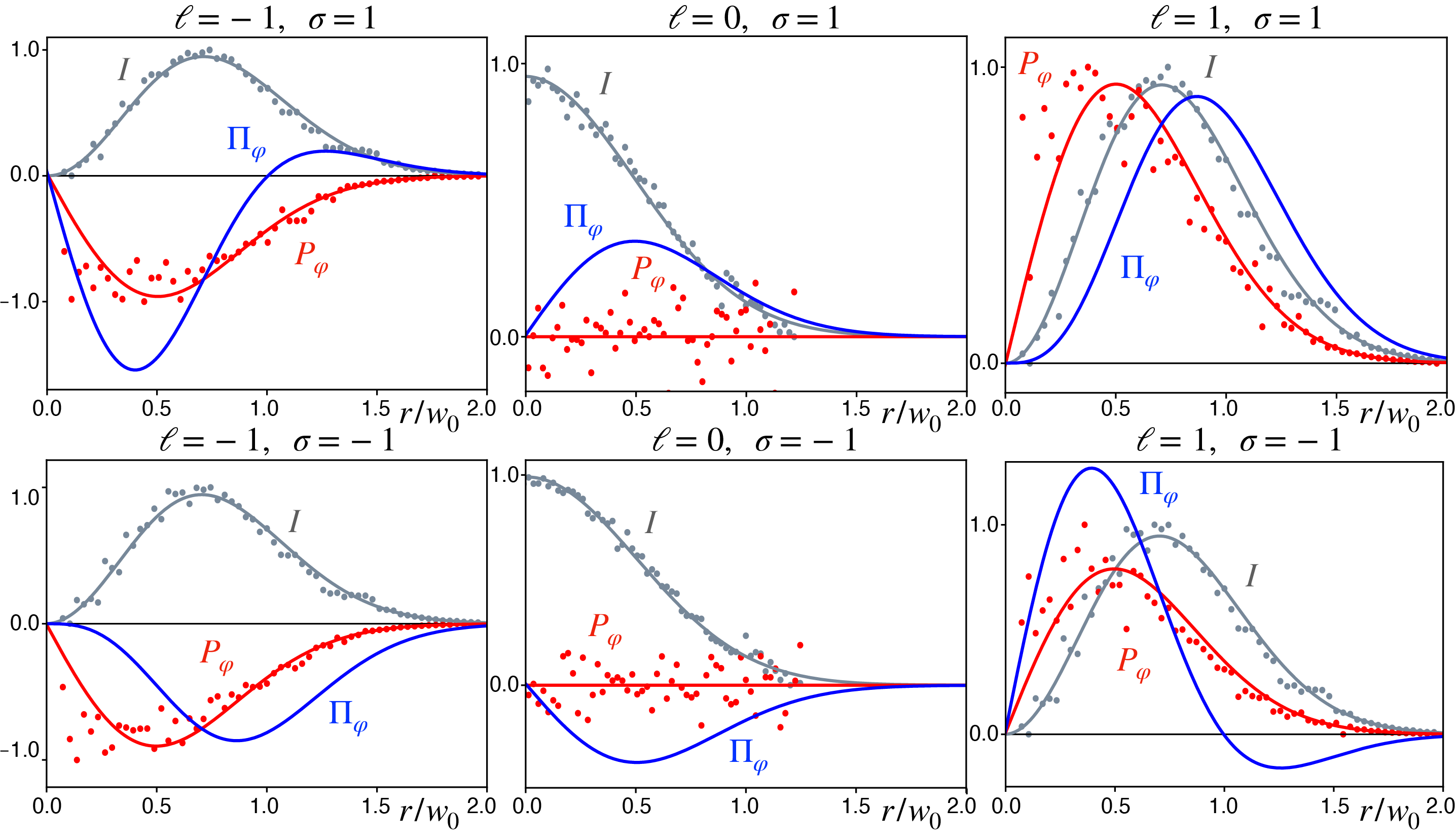}
\caption{Experimental (symbols) vs. theoretical (curves) results for the radial distributions of the intensity and azimuthal component of the momentum density in LG beams with $\ell = 0, \pm 1$ and circular polarization $\sigma = \pm 1$ in the focal $z=0$ plane (plotted in arbitrary units). The experimental results are obtained using the Shack-Hartmann wavefront sensor, as explained in the text and Fig.~\ref{fig2}. \red{These are consistent with the $\sigma$-independent canonical momentum density, Eqs.~(\ref{LG}) and (\ref{P_phi}), rather than with the $\sigma$-dependent Poynting vector (\ref{Poynting_phi})} (see also Fig.~\ref{fig1}).}
\label{fig3}
\end{center}
\end{figure}

The data analysis was performed as follows. The displacements of the centroids of each spot in the spotfield generated by the LG beam were measured with respect to the corresponding spots in the spotfield generated by the reference. The Cartesian coordinates of these displacements were used to obtain the azimuthal component of the displacement \cite{Ghosh2023}. Then, the azimuthal displacement of the $i$th spot centroid, $\Delta\varphi_i$, is divided by the focal length $f_m$ of each microlens and multiplied by the corresponding intensity $I_i$ of the spot. The resulting quantity $P_{\varphi\, i}^{\rm exp} = I_i \Delta\varphi_i / f_m$ provides the experimentally measured azimuthal component of the momentum density, which is in agreement with the canonical momentum (\ref{P_phi}) $P_{\varphi} \propto I \ell/ kr$ (see Fig.~\ref{fig3}). 
We collected several frames of the same spotfield image. For each such frame, the measured intensity and canonical momentum density are grouped into bins corresponding to segments of radii $r_j$. The average over all the frames of intensity $\langle I \rangle_j$  and canonical momentum density  $\langle P_{\varphi}^{\rm exp} \rangle_j$ are obtained for each $r_j$. 

The results of the measurements are shown in Fig.~\ref{fig3}. The grey symbols correspond to the measured intensity $\langle I \rangle_j$ as a function of the radius $r$, while the solid grey curves correspond to the theoretical dependences $|\Psi_\ell (r) |^2$, Eq.~(\ref{LG}), optimally fitted to the experimental measurements. The red and blue curves correspond to the theoretical distributions of the azimuthal components of the canonical momentum density $P_\varphi (r)$ and Poynting momentum $\Pi_\varphi (r)$, Eqs.~(\ref{P_phi}) and (\ref{Poynting_phi}), respectively. Evidently, the measured azimuthal momentum densities (red symbols) are consistent with the canonical rather than Poynting momentum. While the azimuthal canonical momentum is independent of the polarization helicity $\sigma$ and does not change its sign across the beam radius, the azimuthal component of the Poynting vector depends on $\sigma$ and alters its sign with $r$. 
Our experimental results also confirm recent theoretical arguments \cite{Zheng2021} that the Shack-Hartmann wavefront sensor measures the weak value of the canonical momentum operator, i.e., the canonical momentum density. 

\blue{We finally note that for the experimental data shown in Fig.~\ref{fig3}, the statistical error in the measured azimuthal momentum density (red points) ranges from $5\times 10^{-3}$ to $6\times 10^{-2}$ across radius $r$. The maximum statistical error is typically observed near the vortex core as the intensities of the detected spots are low in these regions, resulting in imprecise detection of the corresponding spot centroids. It can also be observed that these data points for vortex beams ($\ell \neq 0$) exhibit some systematic errors around $r/w_0=0.5$ and deviate from the theoretical red curves. These radii correspond to high intensity gradients (as seen from the grey curves), where systematic errors may appear due to cross-talks between neighbouring microlenses (see supplementary section of \cite{Ghosh2023} for more details).}

\section{Propagational dynamics of truncated vortex beams}

Another method to reveal the local propagational evolution of vortex beams is to consider diffraction of such beams after a suitable (e.g., knife-edge) aperture. This method has been applied in optics \cite{Arlt2003, Hamazaki2006,Cui2012,Bekshaev2014} and later for electron vortex beams \cite{Guzzinati2013,Schattschneider2014}. However, the optical case was considered for linearly polarized beams, and the observed evolution was attributed to the Poynting vector. We argue that in fact this evolution is determined by the canonical momentum density. 

\begin{figure}[!t]
\begin{center}
\includegraphics[width=0.8\linewidth]{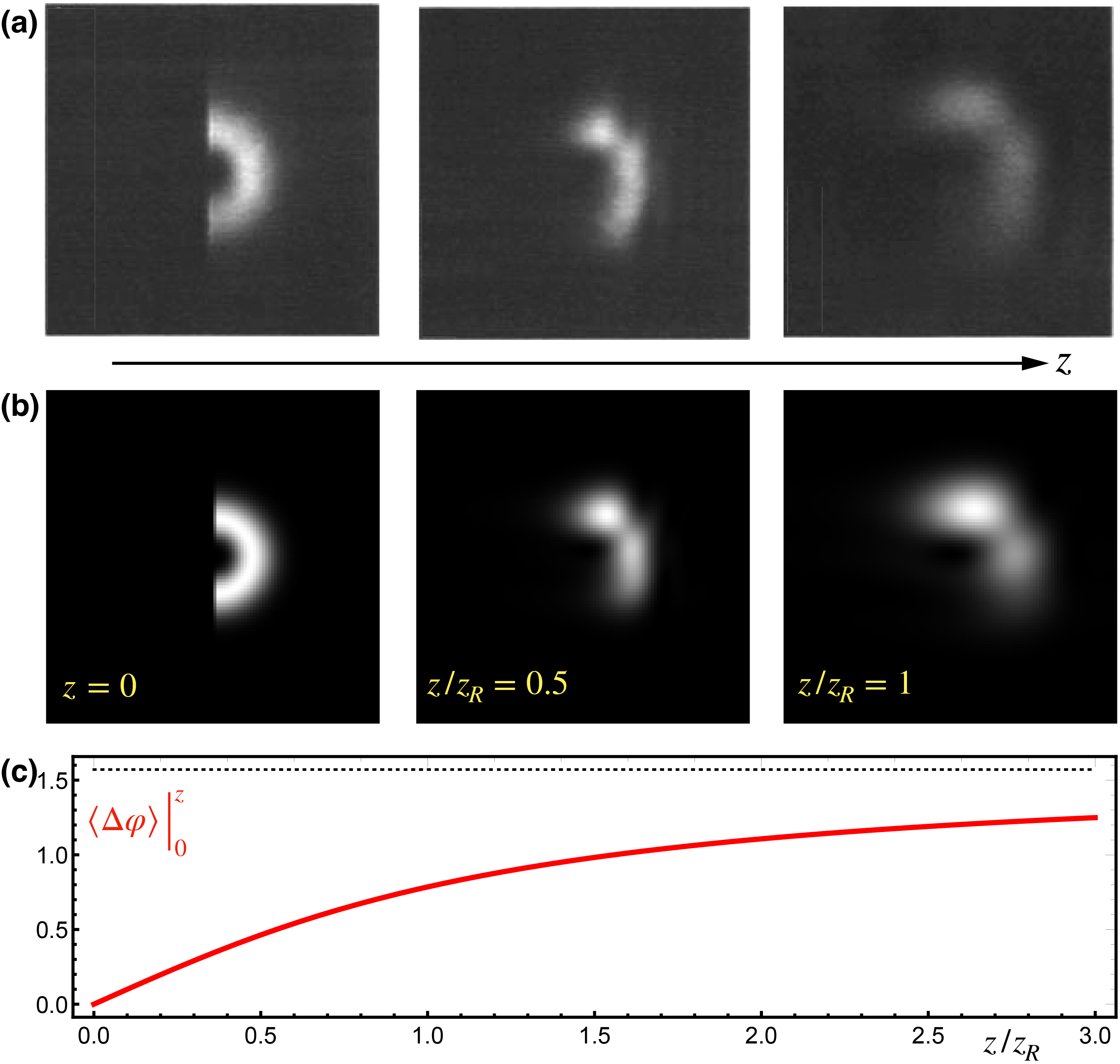}
\caption{{\bf (a)} Experimental pictures from \cite{Arlt2003} of the transverse intensity distributions of an LG beam (\ref{LG}) with $\ell=2$ diffracted at the knife-edge aperture (located in the focal $z=0$ plane) at different distances $z$. {\bf (b)} Numerical calculations of this knife-edge diffraction for the LG beam with $\ell=2$ and $kw_0 = 20$. {\bf (c)} The integral angle of rotation from the $z=0$ plane, given by the Gouy phase $\Phi_G = \arctan (z/z_R)$, Eq.~ (\ref{angle}).}
\label{fig4}
\end{center}
\end{figure}

We first recall the known facts. A vortex beam truncated by a knife edge or a similar aperture, usually placed in the focal plane, experiences an azimuthal rotation upon the $z$-propagation, as shown in Fig.~\ref{fig4}. The direction of this rotation is determined by ${\rm sgn}(\ell)$ and its magnitude for LG beams is given by the Gouy phase $\Phi_G (z)$ \cite{Arlt2003, Hamazaki2006, Cui2012, Guzzinati2013}. The latter fact hints at the connection between the azimuthal rotation and diffraction, although similar rotations have also been observed for non-diffracting Bessel beams \cite{Arlt2003}. However, despite the general association with the Poynting vector \cite{Arlt2003, Hamazaki2006}, an accurate quantitative relation between this rotational evolution and the momentum density in vortex beams remains somewhat elusive. Even for linearly polarized beams, Berry and McDonald remarked that the rotation of the streamlines (trajectories) determined by the momentum density in LG beams ``is unrelated to the Gouy phase'' \cite{Berry2008}. 

\subsection{Laguerre-Gaussian beams}

To characterize the rotational evolution of truncated vortex beams within the optical-current paradigm, we employ the method used in \cite{Schattschneider2014} for electron vortex beams. Namely, assuming that the local velocity of the propagation in structured optical field is given by ${\bf v} = c^2{\bf P}/W$ \cite{Lekner2002, Bliokh2013NJP_II}, we introduce the local angular velocity of the beam rotation with respect to the $z$-coordinate (rather than time):
\begin{equation}
\label{angular}
\Omega = \frac{v_\varphi}{c r} = \frac{cP_\varphi}{rW} = \frac{\ell}{kr^2}  \,.
\end{equation}
Note that such radial distribution of the azimuthal velocity $v_\varphi \propto 1/r$ corresponds to  `irrotational' hydrodynamical vortices, with $({\bm \nabla} \times {\bf v})_z = 0$ and constant normalized angular-momentum density $l_z \propto r v_\varphi$ \cite{Bekshaev2004}. To characterize the global rotation of the beam, we calculate the mean value of this angular velocity akin to a quantum-mechanical expectation value:
\begin{equation}
\label{angular_mean}
\langle \Omega \rangle = \frac{\int_0^\infty \Omega |\Psi_\ell |^2 r dr}{\int_0^\infty |\Psi_\ell |^2 r dr} = \frac{2 {\rm sgn}(\ell)}{kw^2} = {\rm sgn}(\ell) \frac{d \Phi_G}{dz}  \,,
\end{equation}
where we used equations for the LG beam (\ref{LG}). Correspondingly, the angle of rotation of the beam with respect to the focal $z=0$ plane is given by
\begin{equation}
\label{angle}
\langle \Delta \varphi \rangle \big\rvert_0^z = \int_0^z \langle \Omega \rangle\, dz = {\rm sgn}(\ell) \Phi_G(z)  \,,
\end{equation}
in exact agreement with observations of \cite{Arlt2003, Hamazaki2006, Cui2012, Guzzinati2013}.

Equations (\ref{angular})--(\ref{angle}) establish the direct relation between the azimuthal momentum density and the Gouy-phase rotation of truncated LG beams. \red{It appears in the integral expectation values rather than in the local momentum density (i.e., individual Bohmian trajectories or weak values).} Remarkably, if we use the Poynting momentum density $\Pi_\varphi$, Eq.~(\ref{Poynting_phi}), instead of the canonical momentum $P_\varphi$, Eq.~(\ref{P_phi}), 	
\begin{equation}
\label{angular_Poynting}
\Omega_\Pi = \frac{c\Pi_\varphi}{rW} =  \frac{\ell}{k r^2} - \frac{\sigma |\ell |}{k r^2} + \frac{2 \sigma}{k w^2} \,  ,
\end{equation}
the final mean results (\ref{angular_mean}) and (\ref{angle}) remain unchanged: 
\begin{equation}
\label{mean_Poynting}
\langle \Omega_\Pi \rangle = \langle \Omega \rangle \,, \quad
\langle \Delta \varphi_\Pi \rangle = \langle \Delta \varphi \rangle \,  .
\end{equation}

Thus, the global rotation cannot discriminate between the Poynting and canonical momentum densities. To do this, one has to measure the {\it local} $r$-dependent rotations in vortex beams. In  \cite{Arlt2003}, following \cite{Padgett1995}, this was done by involving the concept of the `radial maximum trajectory' $ r = r_{\rm max}(z)$ based on the $z$-evolution of the radial maximum of the intensity distribution and by measuring rotations of the intensity rings with different $r_{\rm max}$ in beams with multiple radial maxima (e.g., Bessel beams). However, this concept of a single radial trajectory is elusive (the momentum density streamlines generate a continuum of trajectories \cite{Berry2008}, Fig.~\ref{fig1}), and we would like to deal with the continuous momentum-density field. 

\subsection{Bessel and other vortex beams}

Note that the expressions (\ref{P_phi}) and (\ref{angular}) for the $\ell$-dependent azimuthal component of the canonical momentum and the corresponding local angular velocity are universal for  vortex beams with the azimuthal dependence $\propto \exp(i\ell\varphi)$. Consider, for example, paraxial non-diffracting Bessel beams \cite{McGloin2005}:
\begin{equation}
\label{Bessel}
\Psi_\ell \propto J_{\ell}(\kappa r) \exp\!\left(i\ell \varphi +ik_z z \right),
\end{equation}
where $J_{\ell}$ is the Bessel function of the first kind, $\kappa \ll k$ is the radial wavevector component, and $k_z = \sqrt{k^2 - \kappa^2}$. The canonical momentum density in paraxial Bessel beams (\ref{Bessel}), in the linear approximation in $\kappa$, equals:
\begin{equation}
\label{P_bessel}
{\bf P} \simeq \frac{1}{c} \left( \frac{\ell}{k r} \bar{\bm \varphi}  + \bar{\bf z} \right) |\Psi_\ell |^2\,.
\end{equation}
%

\begin{figure}[!t]
\begin{center}
\includegraphics[width=0.8\linewidth]{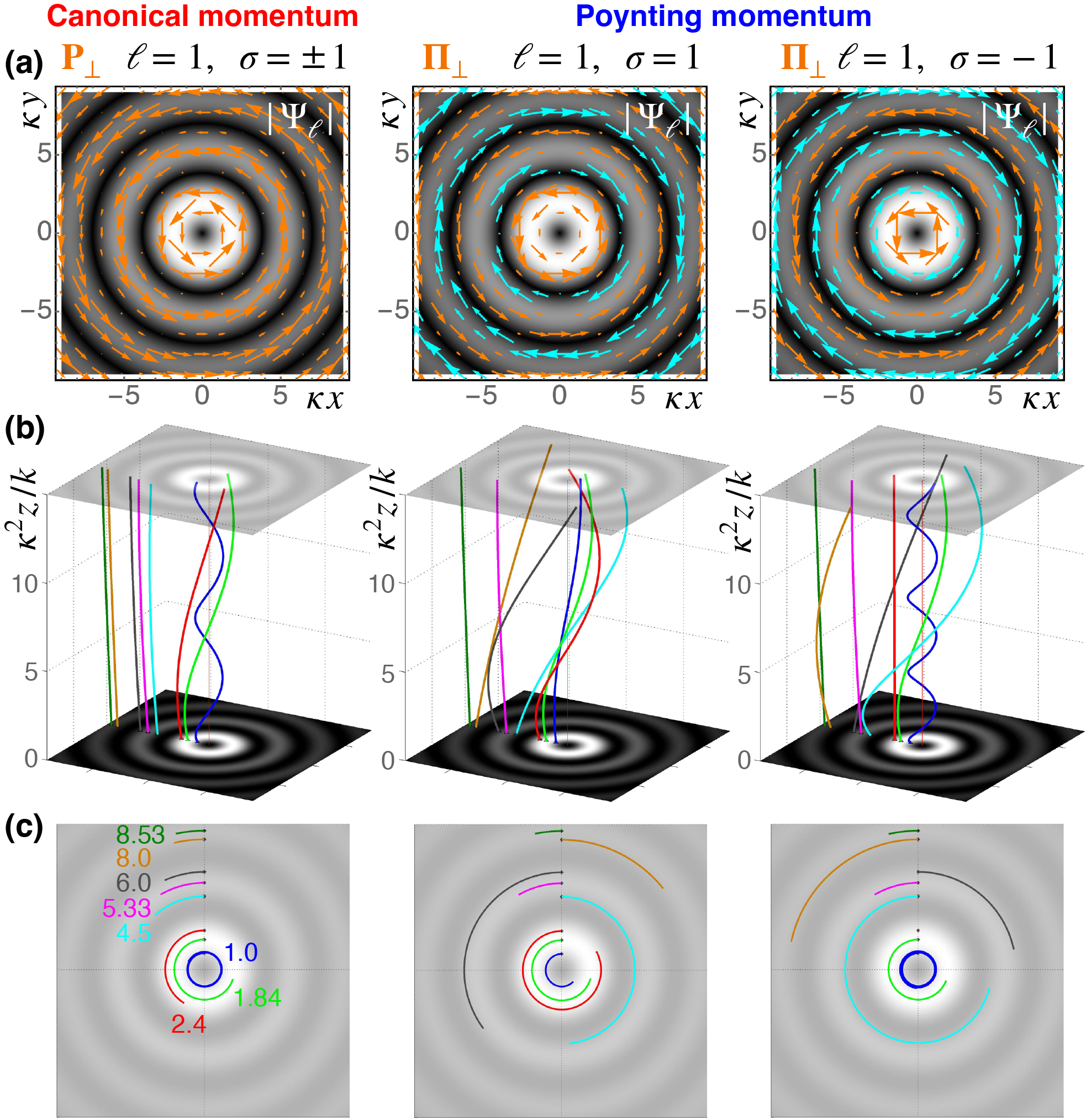}
\caption{The same as in Fig.~\ref{fig1} but for paraxial Bessel beams, Eqs.~(\ref{Bessel})--(\ref{Poynting_Bessel}). The coloured numbers in the lower left panel indicate the values of $\kappa r_0$.}
\label{fig5}
\end{center}
\end{figure}

In contrast, the azimuthal component of the Poynting vector differs considerably for different types of vortex beams. For paraxial Bessel beams, we have $E_z \propto -i\sigma (\kappa/\sqrt{2}k) J_{\ell+\sigma} (\kappa r) e^{i(\ell+\sigma)\varphi}$ \cite{Bliokh2010}, which results in the Poynting momentum \cite{Volke2002}:
\begin{equation}
\label{Poynting_Bessel}
{\bm \Pi} \simeq \frac{1}{c}  \left[ \left( \frac{\ell}{kr}  - \frac{\sigma}{k J_{\ell} (\kappa r)} \frac{d J_{\ell} (\kappa r)}{dr} \right)\!\bar{\bm \varphi} + 
\bar{\bf z} \right] |\Psi_\ell |^2
\propto \frac{\kappa}{ck}\, J_{\ell} (\kappa r) J_{\ell+\sigma} (\kappa r)  \,.
\end{equation}
Here the last expression is entirely similar to the azimuthal probability current in the Dirac-electron Bessel beams \cite{Bliokh2011PRL}. 

Figure~\ref{fig5} shows the transverse distributions of the canonical and Poynting momentum and energy densities, as well as the streamlines of the canonical and Poynting currents for circularly-polarized Bessel beams with $\ell=1$ and $\sigma=\pm 1$. All the streamlines are spirals in this case because of the absence of the radial momentum components (i.e., the absence of diffraction), see Appendix A. Similar to the LG case, the canonical and Poynting streamlines coincide at the radial intensity maxima, i.e., for $d J_\ell (\kappa r)/ d r =0$. Furthermore, akin to the LG-beam case, the azimuthal component of the Poynting current has $r$-regions with positive and negative $\Pi_\varphi$. Thus, if the Poynting momentum density were to determine the $z$-evolution of truncated vortex beams, we could observe local negative rotations in certain $r$-ranges. Moreover, such rotations would strongly depend on the circular polarization $\sigma$. Unfortunately, the experiments \cite{Arlt2003, Hamazaki2006, Cui2012} were performed for linear polarizations, but it is well known that the Kirchhoff-Fresnel integral determining diffraction of paraxial fields \cite{Ananev} does not depend on the polarization. Therefore, we can conclude that the local ($r$-dependent) diffraction evolution of truncated vortex beams cannot be described by the Poynting momentum density. 

\begin{figure}[!t]
\begin{center}
\includegraphics[width=0.9\linewidth]{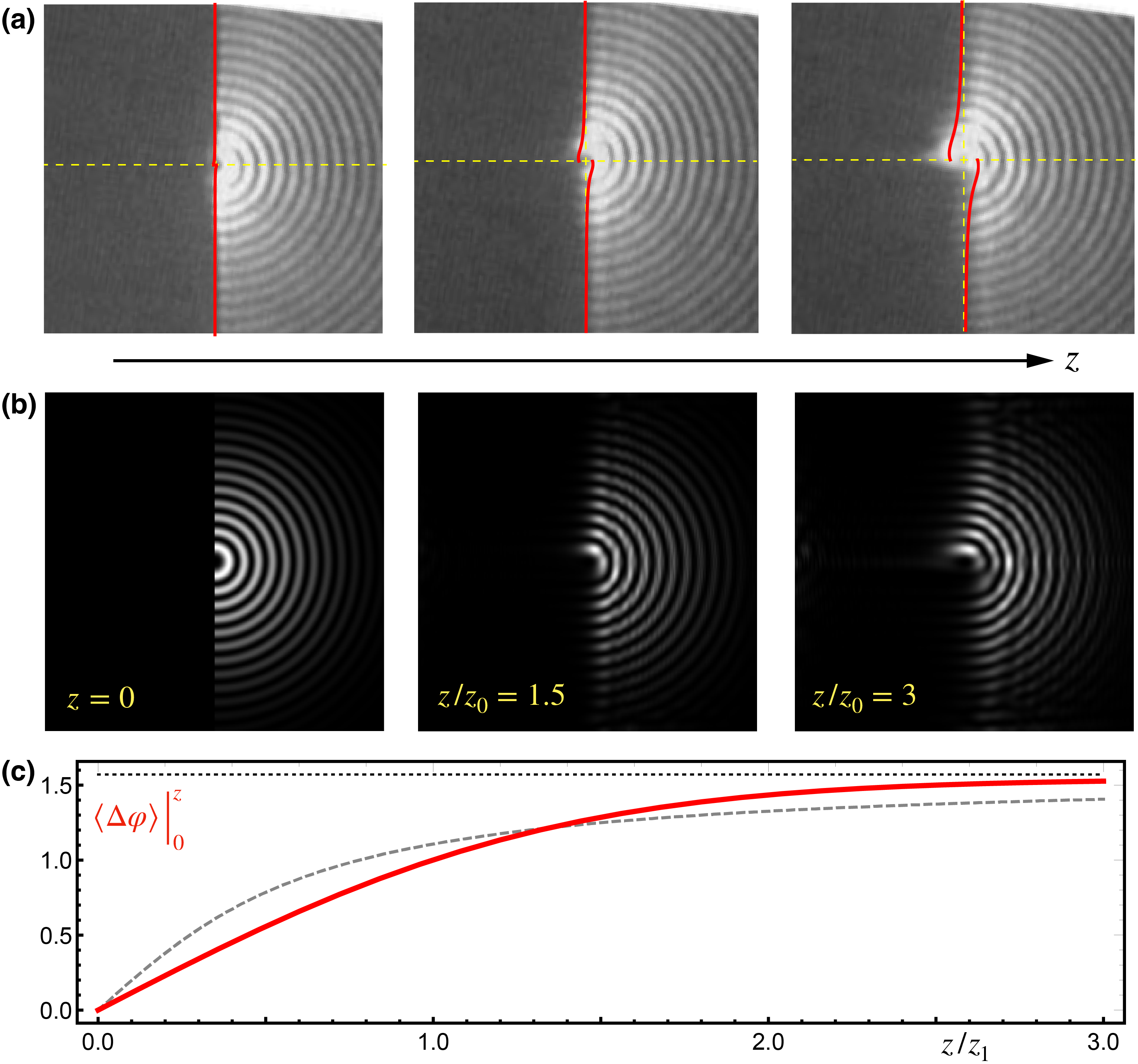}
\caption{{\bf (a)} Experimental pictures from \cite{Arlt2003} of the transverse intensity distributions of a Bessel beam with $\ell=2$ diffracted at the knife-edge aperture (located in the $z=0$ plane) at different distances $z$. Theoretical fits of the knife-edge profile distorted by the $r$-dependent azimuthal rotation (\ref{angle_local}) is shown in red. {\bf (b)} Numerical calculations of the similar knife-edge diffraction of the Bessel-Gauss beam (\ref{BG}) with $\ell =1$, $\kappa/k = 0.2$, and $k w_0 = 150$. The characteristic $z$-scale of the rotation of the inner maximum is $z_0=2k/\kappa^2$. {\bf (c)} Numerically calculated integral angle of rotation from the $z=0$ plane, Eq.~ (\ref{angle_BG}), in the Bessel-Gauss beam (\ref{BG}) with $\ell =1$, $\kappa/k = 0.2$, and $k w_0 = 100$. The characteristic $z$-scale of this integral rotation is $z_1 = kw_0/2\kappa \gg z_0$. Although the $z$-dependence of this rotation differs from the Gouy-phase $\arctan$ function \red{(for reference, an example of such function, $\arctan (2z/z_1)$, is plotted by a dashed line)}, it asymptotically approaches the same $\pi/2$ limit (dotted line), in agreement with the universal Eq.~(\ref{angle_total}).}
\label{fig6}
\end{center}
\end{figure}

Although the local $r$-dependent angular velocity (\ref{angular}) determined by the canonical momentum is independent of the form of the vortex beam $\Psi_\ell$, the resulting $z$-evolution can depend on it. Hence, we can write the {\it local} angle of the truncated vortex rotation as  
\begin{equation}
\label{angle_local}
\Delta \varphi (r,z) = 
\frac{\ell}{r^2} F(z)  \,,
\end{equation}
where $F(z)$ is a function depending on the beam profile $\Psi_\ell$, which provides a $z$-scaling of the same universal $r$-dependent rotation. Figure~\ref{fig6}(a) shows that the suitably scaled deformation of the knife-edge line described by Eq.~(\ref{angle_local}) agrees very well with the experimentally-observed \cite{Arlt2003} rotation of multiple truncated rings in a Bessel beam.

The theoretical model of \cite{Arlt2003}, based on the `radial maximum trajectory', used the expression (\ref{angle_local}) with $F(z) = w^2(z) \Phi_G (z) /2$ involving the width and Gouy phase for the LG beam (\ref{LG}). Apparently, for Bessel or other beams, this function should have a different form. Since Bessel beams are non-normalizable (the integral $\int_0^\infty J_\ell^2 (\kappa r)\, r\,dr$ diverges), let us now consider normalizable and diffracting Bessel-Gauss beams \cite{Borghi2001}:
\begin{equation}
\label{BG}
\Psi_\ell \propto \frac{z_R}{z_R +i z}\, J_\ell \!\left( \frac{z_R \kappa r}{z_R+iz} \right) \exp\!\left[-\frac{k^2 r^2 +i \kappa^2 z z_R }{2k(z_R + i z)} +i\ell \varphi +ikz \right].
\end{equation}
In the case of $\kappa  w_0 = \kappa \sqrt{2 z_R /k} \gg 1$ (which we consider hereafter), such beams contain multiple Bessel-function intensity rings within the bright central part of the Gaussian envelope. 
Figure~\ref{fig6}(b) shows numerically calculated diffraction of such a beam on a knife-edge aperture. One can clearly see the $r$-dependent rotation, described by Eq.~(\ref{angle_local}), in different truncated rings. Remarkably, the maximum rotation angle from $z=0$ to $z=\infty$ in the inner ring is still $\pi/2$, as for the Gouy phase in the LG beams. 

To understand this result we calculate the integral (over all rings) rotation angle for the Bessel-Gauss beam (\ref{BG}) similar to Eqs.~(\ref{angular})--(\ref{angle}):
\begin{equation}
\label{angle_BG}
\langle \Delta \varphi \rangle \big\rvert_0^z = \frac{\ell}{k} \int_0^z \frac{\int_0^\infty |\Psi_\ell |^2 r^{-1} dr}{\int_0^\infty |\Psi_\ell |^2 r dr} \, dz  \,,
\end{equation}
The $z$-dependence of this rotation angle is shown in Fig.~\ref{fig6}(c). It resembles the arctan shape of the Gouy phase, but is not equivalent to it. Moreover, the characteristic $z$-scale of this dependence is $k w_0 /2 \kappa \ll z_R$. 

We have found analytically that the total integral rotation for the whole range $z\in (-\infty, \infty)$ equals
\begin{equation}
\label{angle_total}
\langle \Delta \varphi \rangle \big\rvert_{-\infty}^{\infty} = \pi\, {\rm sgn}(\ell) 
\end{equation}
for {\it any} paraxial vortex beams of the form $\Psi_\ell = \psi (r,z) e^{i\ell\varphi}$ (see Appendix B). 
This result is more fundamental than the Gouy-phase Eqs.~(\ref{angular_mean}) and (\ref{angle}) limited to the LG beams. The fundamental global rotation (\ref{angle_total}) can be understood within the geometrical-optics ray picture. Indeed, the far-field $z=-\infty$ and $z=\infty$ regions can be connected by straight geometrical-optics rays \cite{Berry2008}, and projection of any straight ray onto the transverse $(x,y)$-plane covers the $\pi$ range in the azimuthal angle $\varphi$, while the direction of the azimuthal-angle evolutions is determined by ${\rm sgn}(\ell)$. 
Thus, the integral relation (\ref{angle_total}) provides the connection between the local canonical-momentum current and the global geometrical-optics ray picture.

\section{On the `supermomentum' and `backflow'}

Here we briefly discuss the canonical and Poynting currents in the context of the `supermomentum' \cite{Berry2009,Barnett2013} and `backflow' \cite{Bracken1994,Berry2010}. Both phenomena are currently attracting significant attention \cite{Afanasev2021,Ivanov2022,Eliezer2020,Daniel2022,BB2022,Ghosh2023} and represent examples of {\it superoscillations} \cite{Berry2019}.
Namely, if the wavefield consists of plane waves (Fourier spectrum) with some component of the wavevectors limited by the range $k_i \in (0,k)$, then the spatial regions where the local phase gradient (i.e., the local wavevector or normalized momentum density $\omega P_i / W$) is larger than $k$ or less than $0$ correspond to the supermomentum and backflow, respectively \cite{Bliokh2013NJP_II}.
In other terms, these conditions correspond to the superluminal or negative local velocity $v_i = c^2 P_i / W$: $v_i > c$ and $v_i < 0$. In the context of vortex beams, we consider these phenomena for the azimuthal momentum component, or angular momentum.  

\subsection{Supermomentum near the vortex core}

For the azimuthal component of the canonical momentum density (\ref{P_phi}) the supermomentum  occurs in the subwavelength vicinity of the vortex centre: 
\begin{equation}
\label{P_super}
\frac{\omega |P_\varphi|}{W} \simeq  \frac{|\ell |}{r}  > k\,, \quad 
r < |\ell | k^{-1} \,.
\end{equation}
Remarkably, the supermomentum can never been achieved for the Poynting momentum density: 
\begin{equation}
\label{Poynting_super}
\frac{\omega |{\bm \Pi}|}{W} \leq k\,,
\end{equation}
%
which follows from the general definitions (\ref{W}) and (\ref{Poynting}). Although the supermomentum near the vortex centre has never been measured in experiments, it has been observed in evanescent waves (where the local phase gradient is larger then $k$) using light-matter interactions with atoms \cite{Huard1978,Matsudo1998}. Moreover, the sharp contrast of this measured supermomentum with the Poynting momentum was emphasized in these studies (see also \cite{Bliokh2013NJP_II,Bliokh2014NC}). 

However, the azimuthal component of the Poynting vector in Eqs.~(\ref{Poynting_phi}) and (\ref{Poynting_Bessel}) for ${\rm sgn}(\sigma\ell) =-1$ exhibits the same $1/r$ divergence (when divided by the intensity) as the canonical momentum and hence contradicts general condition (\ref{Poynting_super}). This means that our paraxial-approximation equations become inapplicable near the vortex center, and the second-order terms become important in this region. 
Indeed, near the vortex core (e.g., for $r\ll w_0$ in LG beams), $\partial \Psi_\ell / \partial r \simeq (|\ell |/ r) \Psi_\ell$, and the longitudinal field component is estimated as $|E_z| \sim (|\ell | /kr) |\Psi_\ell |$. Then, its contribution becomes of the order of the main field, $|E_z| \sim |\Psi_\ell |$, for $r \lesssim |\ell | k^{-1}$, i.e., exactly in the supermomentum area (\ref{P_super}). 

Importantly, this longitudinal-field contribution includes spin-orbit interaction effects \cite{Bliokh2010, Bliokh2015NP}, and all the beam properties become strongly $\sigma$-dependent in the supermomentum area. To take these effects into account, we calculate the azimuthal components of the canonical and Poynting momentum, and also the energy density, for circularly-polarized Bessel beams without using paraxial approximation. In doing so, we use the exact Bessel-beam solutions described in \cite{Bliokh2010}. These are helicity eigenmodes, i.e., represent a superposition of plane waves with different wave vectors and the same circular polarization (helicity) $\sigma=\pm 1$. Owing to this, the magnetic field exactly satisfies Eq.~(\ref{H}): ${\bf H} = -i\sigma {\bf E}$, whereas the Cartesian components of the electric field can be written as:
\begin{align}
\label{Bessel_exact}
E_x & \propto \frac{1}{\sqrt{2}} \left[ a J_\ell (\kappa r) +b e^{2i\sigma\varphi} J_{\ell+2\sigma} (\kappa r) \right] e^{i\ell\varphi + i k_z z}\,, \nonumber \\
E_y & \propto \frac{i\sigma}{\sqrt{2}} \left[ a J_\ell (\kappa r) - b e^{2i\sigma\varphi} J_{\ell+2\sigma} (\kappa r) \right] e^{i\ell\varphi + i k_z z}\,, \\
E_z & \propto -i \sigma \sqrt{2ab}\, e^{i\sigma\varphi} J_{\ell+\sigma} (\kappa r) e^{i\ell\varphi + i k_z z}\,, \nonumber 
\end{align}
where $a=(k+k_z)/2k$, $b=(k-k_z)/2k$, and $2 \sqrt{ab} = \kappa/k$. 
Substituting these fields into general Eqs.~(\ref{W})--(\ref{P}) and using the radial field component $E_r \propto (e^{i\sigma\varphi} /\sqrt{2})\left[a J_\ell (\kappa r) +b J_{\ell+2\sigma} (\kappa r)\right]e^{i\ell\varphi + i k_z z}$, we derive the azimuthal momentum and energy density distributions:
\begin{equation}
\label{P_exact}
P_\varphi \propto \frac{1}{ckr} \left[ a^2 \ell  J_\ell^2 (\kappa r) + b^2 (\ell+2\sigma) J_{\ell+2\sigma}^2 (\kappa r) + 2ab (\ell +\sigma) J_{\ell+\sigma}^2 \right] ,
\end{equation}
\begin{equation}
\label{Poynting_exact}
\Pi_\varphi \propto \frac{\kappa}{ck} \left[a J_{\ell} (\kappa r) + b J_{\ell +2\sigma} (\kappa r)\right]J_{\ell+\sigma} (\kappa r)  \,,
\end{equation}
\begin{equation}
\label{W_exact}
W \propto \left[ a^2 J_\ell^2 (\kappa r) + b^2 J_{\ell+2\sigma}^2 (\kappa r) + 2ab J_{\ell+\sigma}^2 \right] .
\end{equation}
The previously used paraxial approximation corresponds to $a\simeq 1$ and $b\simeq 0$. Note that the spin-orbit effects result in the $\sigma$-dependent terms (proportional to $b$ and $b^2$) in the canonical momentum and energy distributions (\ref{P_exact}) and (\ref{W_exact}).

\begin{figure}[!t]
\begin{center}
\includegraphics[width=\linewidth]{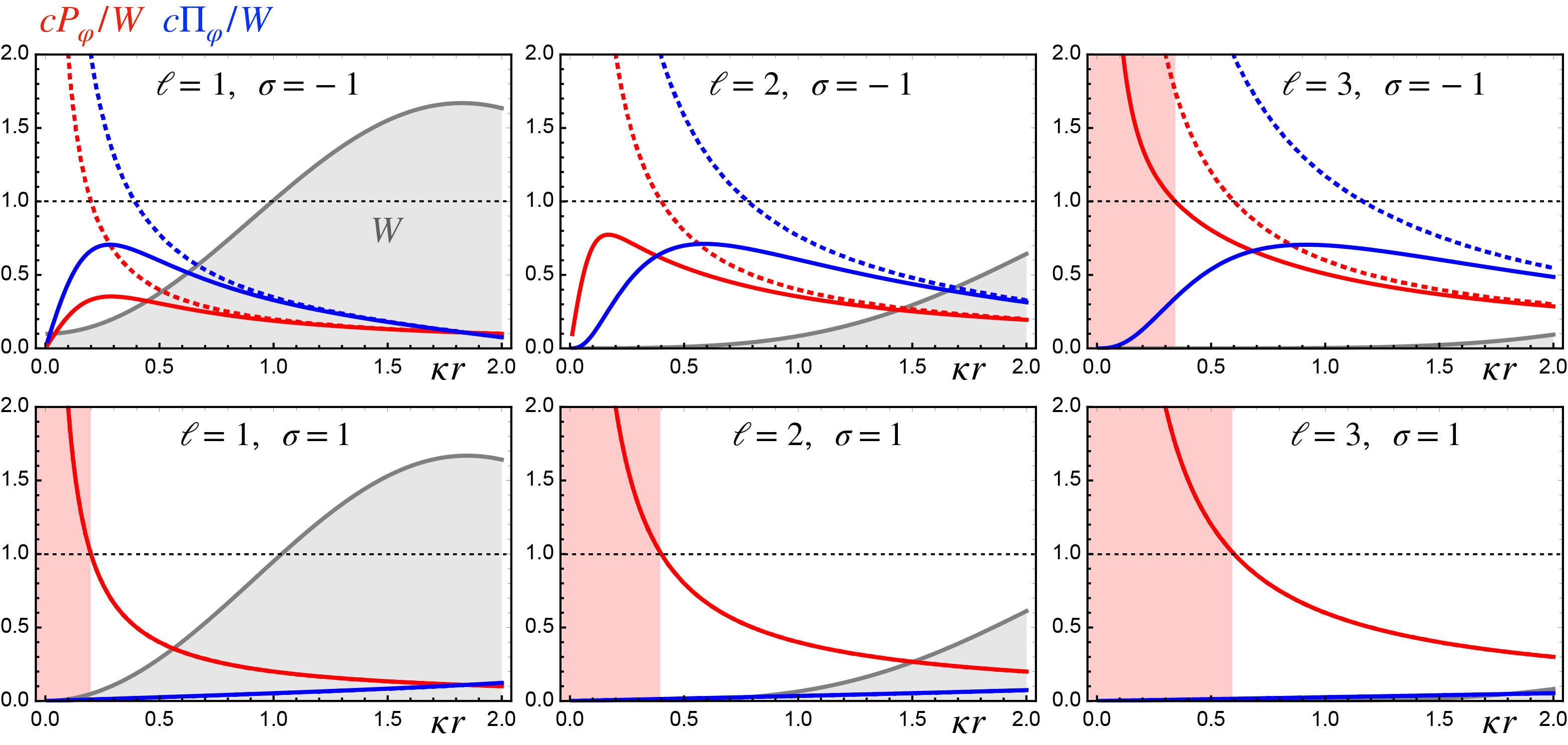}
\caption{Radial dependences of the normalized azimuthal components of the canonical and Poynting momentum densities in circularly-polarized Bessel beams with $\kappa/k =0.2$, $\ell=1,2,3$ and $\sigma =\pm 1$. Exact expressions are given by Eqs.~(\ref{P_exact})--(\ref{W_exact}), while paraxial approximation, shown by dashed curves, corresponds to $a\simeq 1$ and $b\simeq 0$. The areas of the canonical supermomentum $c|P_\varphi |/W >1$ are highlighted by pink. The grey profiles indicate the energy-density distributions $W(r)$. The simultaneous sign flip $(\ell,\sigma) \to (-\ell,-\sigma)$ reverses the azimuthal momentum components: 
$P_\varphi \to - P_\varphi$, $\Pi_\varphi \to - \Pi_\varphi$}.
\label{fig7}
\end{center}
\end{figure}

Figure~\ref{fig7} shows the normalized values $cP_\varphi/W$ and $c\Pi_\varphi /W$ as functions of $r$, both exact and in the paraxial approximation, for Bessel beams with $\ell =1,2,3$ and $\sigma=\pm 1$. One can see drastic difference between the paraxial and exact curves for ${\rm sgn}(\sigma\ell) =-1$. In agreement with the general relation (\ref{Poynting_super}), the exact Poynting vector never corresponds to supermomentum. Furthermore, the exact canonical momentum does not produce supermomentum in the case of ${\rm sgn}(\sigma\ell) =-1$ and $|\ell| \leq 2$. \red{Under these conditions $W(0)\neq 0$, and, hence, the azimuthal supermomentum can be achieved only in the case when the energy density (including all the field components) exactly vanish in the vortex center. One can say that it is the vanishing energy density, rather than the anomalous momentum density, that provides for the supermomentum.}

%
%


\subsection{Azimuthal backflow}

Although, the backflow is usually considered for Cartesian components of the wave current, recently it was shown that it can be equally considered for the local azimuthal phase gradient \cite{Ghosh2023}. In this case, one has to consider the expansion of the wavefield in the orbital-angular-momentum (vortex) eigenmodes instead of plane waves. In particular, if the fields represents a superposition of the modes with $\ell > 0$, then the negative local azimuthal phase gradient corresponds to the azimuthal backflow. 

In this context, we note that the azimuthal component of the Poynting vector in a circularly polarized vortex beam with $\ell > 0$ and $\sigma = -1$, Eqs.~(\ref{Poynting_phi}) and (\ref{Poynting_Bessel}), becomes negative in certain $r$-regions, see Figs.~\ref{fig1} and \ref{fig5}. However, one should not interpret it as an azimuthal backflow \cite{BB2022}. Such vortex beams are orbital-angular-momentum eigenmodes with fixed eigenvalues $\ell$ (in the paraxial regime one can consider the orbital and spin parts of the angular momentum independently \cite{Allen1999,Bliokh2015PR,Bliokh2015NP}), while superoscillations are interference phenomena which appear only in superpositions of several momentum or angular-momentum eigenmodes \cite{Berry2019}. Moreover, as we have shown, the Poynting momentum density does not correspond to the local phase gradient, or wavevector, or propagation direction of light, and the actual optical current corresponds to the canonical momentum density. Its azimuthal component is always positive in a vortex beam with fixed $\ell >0$, and the interference of at least two vortex beams is needed to achieve the true azimuthal backflow \cite{Ghosh2023}. 

\section{Conclusions}

In this work, we aimed to overview the roles of the canonical and Poynting momentum densities (currents) in the propagation and diffraction of structured monochromatic optical beams. In doing so, we considered vortex beams of different kinds. While the Poynting vector is determined by the well-known vector product of the electric and magnetic fields, the canonical momentum density is associated with the wave {\it phase gradient} ({\it local wavevector}) averaged over all the field components and multiplied by the wave intensity. In this manner, the canonical momentum provides direct link to the quantum-mechanical de Broglie momentum and probability current. It is well established in other works that local {\it optomechanical} momentum effects (i.e., optical forces on small particles or momentum exchange with atoms) are determined by the canonical momentum density rather than the Poynting vector \cite{Ashkin1983,Bliokh2014NC,Bliokh2015PR,Marques2014,Leader2016}. Furthermore,  {\it quantum weak measurements} of the local momentum also result in the canonical momentum density \cite{Berry2009,Kocsis2011,Bliokh2013NJP_II}. In this work, we have considered the canonical-Poynting dilemma in the context of purely optical {\it propagation and diffraction} phenomena.  

In the paraxial regime, assuming uniformly polarized light, the canonical current is independent of the polarization, while the Poynting vector depends on the degree of circular polarization, i.e., spin. 
Previous optical approaches to the optical currents in the propagation and diffraction of structured light \cite{Padgett1995,Leach2006,Berry2008,Arlt2003, Cui2012} associated it with the Poynting vector. However, these works dealt with linearly-polarized (zero-spin) beams where the canonical and Poynting currents coincide with each other. We have revisited these approaches using circularly polarized beams and found that it is the canonical momentum density rather than the Poynting vector which should be associated with the observable phenomena in the propagation of light. 

First, we have performed experimental measurements of the transverse momentum density in vortex LG beams via Shack-Hartmann wavefront sensor. This yielded polarization-independent results consistent with the canonical current rather than the Poynting vector. Second, we have analyzed diffraction of different kinds of vortex beams at a knife-edge aperture placed in the focal plane. The truncated beams experience a $z$-dependent rotation which depends on the sign of the vortex charge $\ell$ and can be associated with the Gouy phase. We have shown that this $z$-dependent rotation can be explained by the optical current properly integrated over the transverse cross-section of the beam. Notably, this integral current is the same in the canonical and Poynting versions; only the local $r$-dependent rotation can discriminate between these. However, since the paraxial diffraction is polarization-independent, we can conclude that it must be locally defined by the canonical rather than Poynting momentum density. Remarkably, we have derived a theoretical relation for the total integral rotation over the whole $z\in (-\infty, \infty)$ range in a circularly-symmetric vortex beam with an arbitrary $r$-profile. The angle of this rotation always equals $\pi\, {\rm sgn}(\ell)$, which provides a global relation between the optical currents, Gouy phase, and the geometrical-optics ray picture. While optical currents describe {\it local} propagation properties of a structured wavefield, geometrical optics rays connect {\it far-field} regions in the field evolution.  

Finally, we have discussed the `supermomentum' \cite{Berry2009,Bliokh2013NJP_II,Barnett2013,Afanasev2021,Ivanov2022} and `backflow' \cite{Bracken1994,Berry2010,Bliokh2013NJP_II,Eliezer2020,Daniel2022,BB2022,Ghosh2023} effects in the canonical and Poynting currents in vortex beams. The former phenomenon means anomalously high momentum density normalized by the energy density, and it appears in subwavelength area near the vortex core. We have shown that it is only possible for the canonical momentum density and strongly depends on the nonparaxial spin-dependent corrections. \red{In this regime, the energy and canonical momentum densities become polarization-dependent.} The latter phenomenon means anomalously negative local optical current, and it should be also associated with the canonical momentum density in interfering modes. While the Poynting vector can have negative azimuthal components in vortex beams with $\ell>0$, it should not be associated with the backflow, because the actual propagation is determined by the canonical momentum density. Interference of at least two vortex modes with different $\ell$ is required for the proper canonical azimuthal backflow \cite{Ghosh2023}. 

We hope that this work sheds light on important peculiarities of optical propagation and the relevant physical characteristics for their description. \red{The canonical momentum density provides a useful theoretical tool for understanding the free-space evolution of monochromatic light fields. It would be important to further explore and extend this framework beyond its current limitations: e.g. for polychromatic optical fields and for structured light in anisotropic media.}

\begin{backmatter}
\bmsection{Acknowledgments}
We are grateful to Miguel A. Alonso
for helpful discussions {and to Filippo Cardano for his support with the Q-plates for our experiments}.
{This work is supported in part by 
the Foundation for Polish Science
co-financed by the European Union under the European Regional Development Fund (POIR.04.04.00-00-3004/17-00) [the FIRST TEAM project ``Spatiotemporal photon correlation measurements for quantum metrology and super-resolution microscopy'' and the TEAM project TEAM/2016-3/29]; 
the National Science Centre, Poland [grant 2022/47/B/ST7/03465];
the International Research Agendas Programme (IRAP) of the Foundation for Polish Science 
co-financed by the European Union under the European Regional Development Fund and Teaming Horizon 2020 programme of the European Commission [the ENSEMBLE3 project MAB/2020/14]; 
and the project of the Minister of Science and Higher Education ``Support for the activities of Centers of Excellence established in Poland under the Horizon 2020 program'' [contract MEiN/2023/DIR/3797].}
\end{backmatter}

\bibliography{References}

\newpage 

\renewcommand{\theequation}{A.\arabic{equation}}
\setcounter{equation}{0}
\section*{Appendix A: Streamlines of the canonical and Poynting currents in LG and Bessel beams}

The streamlines $(r(z),\varphi(z))$ of a vector field ${\bf V}({\bf r})$  are solutions of differential equations
\begin{equation}
\label{streamlines}
\frac{dr}{dz} = \frac{V_r}{V_z}\,,\quad
r\frac{d\varphi}{dz} = \frac{V_\varphi}{V_z}\,.
\end{equation}
Substituting here the canonical momentum density (\ref{P_phi}) in LG beams, we obtain 
\begin{equation}
\label{streamlines_LG}
\frac{dr}{dz} = \frac{rz}{z^2 + z_R^2}\,,\quad
\frac{d\varphi}{dz} = \frac{\ell}{kr^2}\,.
\end{equation}
Solutions of these equations yield a family of `Bohmian trajectories'
\begin{equation}
\label{trajectories_LG}
r(z) = r_0 \sqrt{1+\frac{z^2}{z_R^2}} = r_0 \frac{w(z)}{w_0} \,,\quad
\varphi (z)  = \varphi_0 + \frac{\ell z_R}{kr_0^2}\arctan \! \left( \frac{z}{z_R}\right) = 
\varphi_0 + \frac{\ell w_0^2}{2r_0^2}\Phi_G (z)\,,
\end{equation}
where $r_0 = r(0)$ and $\varphi_0 = \varphi (0)$. 

In turn, for the Poynting momentum density (\ref{Poynting_phi}), the azimuthal angle equations are modified as
\begin{equation}
\label{streamlines_LG_Poynting}
\frac{d\varphi_\Pi}{dz} = \frac{\ell - \sigma |\ell |}{kr^2} + \frac{\sigma z_R}{z^2+z_R^2}\,,
\end{equation}
\begin{equation}
\label{trajectories_LG_Poynting}
\varphi_\Pi (z)  = 
\varphi_0 + \left[ (\ell - \sigma |\ell |) \frac{w_0^2}{2r_0^2} +\sigma \right]\!\Phi_G (z)\,.
\end{equation}
Examples of the canonical and Poynting streamlines (\ref{trajectories_LG}) and (\ref{trajectories_LG_Poynting}) in LG beams are shown in Fig.~\ref{fig1}(b,c). The canonical and Poynting streamlines coincide at the radial intensity maximum, i.e., for $r_0 = \sqrt{|\ell|/2} w_0$, where $\partial W/\partial r |_{z=0}=0$.

For Bessel beams (\ref{Bessel})--(\ref{Poynting_Bessel}), the radial evolution is trivial: $dr/dz =0$, $r(z) = r_0$, while the azimuthal equations for the canonical and Poynting streamlines yield:
\begin{equation}
\label{streamlines_Bessel}
\frac{d\varphi}{dz} = \frac{\ell}{kr_0^2}\,, \quad
\varphi (z)  = \varphi_0 + \frac{\ell}{kr_0^2}\,z \,,
\end{equation}
\begin{equation}
\label{streamlines_Poynting,Bessel}
\frac{d\varphi_\Pi}{dz} = \frac{\kappa }{kr_0} \frac{J_{\ell+\sigma} (\kappa r_0)}{J_{\ell} (\kappa r_0)} \,, \quad
\varphi_\Pi (z)  = \varphi_0 + \frac{\kappa }{kr_0} \frac{J_{\ell+\sigma} (\kappa r_0)}{J_{\ell} (\kappa r_0)}\, z \,.
\end{equation}
Examples of these spiral streamlines are shown in Fig.~\ref{fig5}(b,c). Akin to the LG case, the canonical and Poynting streamlines coincide at the radial intensity maxima, i.e., when $d J_\ell (\kappa r) / dr =0$. This follows from the relation $\kappa J_{\ell+\sigma} (\kappa r) = (\ell/r) J_\ell (\kappa r) - \sigma d J_\ell (\kappa r) / dr$. 

\newpage 

\renewcommand{\theequation}{B.\arabic{equation}}
\setcounter{equation}{0}
\section*{Appendix B: Integral rotation angle for arbitrary vortex beams}

Here we calculate the integral rotation angle over the whole range $z\in (-\infty, \infty)$ for an arbitrary paraxial vortex beam $\Psi_\ell = \psi (r,z) e^{i\ell \varphi}$:
\begin{equation}
\label{B1}
\langle \Delta \varphi \rangle \big\rvert_{-\infty}^\infty 
= \int_{-\infty}^\infty \frac{\int_0^\infty \Omega |\Psi_\ell |^2 r dr}{\int_0^\infty |\Psi_\ell |^2 r dr} \, dz  
= \frac{\ell}{k} \int_{-\infty}^\infty \frac{\int_0^\infty |\Psi_\ell |^2 r^{-1} dr}{\int_0^\infty |\Psi_\ell |^2 r dr} \, dz\,,
\end{equation}
which is based on the universal form of the local angular velocity and azimuthal components of the canonical momentum density, Eq.~(\ref{angular}).

To evaluate this expression, we represent the real-space wave function $\Psi_\ell$ by the plane-wave Fourier integral in cylindrical coordinates:
\begin{equation}
\label{B2}
\Psi(r,\varphi,z)=\int_0^{\infty} d\kappa\,  \int_0^{2\pi} d\phi\, \kappa\, \tilde{\Psi}(\kappa)\,e^{i\kappa r \cos(\phi - \varphi)+i\ell\phi +ik_z z} \,,
\end{equation}
where $(\kappa,\phi)$ are polar coordinates in the $(k_x,k_y)$ plane, $k_z = \sqrt{k^2-\kappa^2}\simeq k - \kappa^2/2k$, and $\tilde{\Psi}(\kappa)$ is a function characterizing the radial beam shape. 
The azimuthal integral in (\ref{B2}) can be evaluated to the Bessel function of the first kind:
\begin{equation}
\label{B3}
\Psi(r,\varphi,z)\propto e^{i\ell\varphi + ikz}\,\int_0^{\infty} d\kappa\,  \kappa \, \tilde{\Psi}(\kappa)J_{\ell}(\kappa r) \,e^{-i \kappa^2 z / 2k}  \,.
\end{equation}
Substituting Eq.~(\ref{B3}) into the denominator of Eq.~(\ref{B1}), we obtain:
\begin{align}
\label{B4}
\int_0^\infty |\Psi(r,z) |^2 r dr & \propto \int_0^\infty dr\, \int_0^\infty  d\kappa \int_0^\infty  d\kappa' \, 
r \kappa \kappa' \tilde{\Psi}^*(\kappa')\tilde{\Psi}(\kappa) J_\ell(\kappa'r) J_\ell(\kappa r) e^{i (\kappa'^2-\kappa^2)z/2k}  \nonumber \\
&=\int_0^\infty  |\tilde{\Psi}(\kappa)|^2\, \kappa \, d\kappa\,.
\end{align}
Here we used the ``closure equation'' for the Bessel functions: $\int_0^\infty J_\ell(\kappa'r) J_\ell(\kappa r)\, r\, dr\,= \kappa^{-1} \delta(\kappa-\kappa')$, where $\delta$ is the Dirac delta-function.

Thus, the denominator in Eq.~(\ref{B1}) is $z$-independent, and the z-integration can be performed in the numerator. Substituting there the Fourier representation (\ref{B3}), akin to Eq.~(\ref{B4}), and using the relation 
\begin{equation}
\int_{-\infty}^{\infty} e^{i(\kappa^2-\kappa'^2)z/2k}\, dz  = 2\pi \delta\!\left(\frac{\kappa^2-\kappa'^2}{2k}\right) = \frac{2\pi k}{\kappa}\delta (\kappa-\kappa')\,, \nonumber
\end{equation}
we derive:
\begin{align}
\label{B5}
\int_{-\infty}^{\infty} dz \int_0^\infty |\Psi(r,z) |^2 r^{-1} dr &\propto 2\pi k \int_0^\infty dr\, \int_0^\infty d\kappa\, 
r^{-1}  \kappa \, |\tilde{\Psi}(\kappa)|^2 J_\ell^2(\kappa r)  \nonumber \\
&= \frac{\pi k}{|\ell|} \int_0^\infty  
|\tilde{\Psi}(\kappa)|^2\,\kappa\, d\kappa\,,
\end{align}
where we used $\int_0^\infty J_\ell^2(\kappa r)\,r^{-1}  \, dr= 1/(2|\ell|)$. 

Finally, substituting Eqs.~(\ref{B4}) and (\ref{B5}) into Eq.~(\ref{B1}), we arrive at Eq.~(\ref{angle_total}):
\begin{equation}
\label{B6}
\langle \Delta \varphi \rangle \big\rvert_{-\infty}^\infty 
= \pi \, {\rm sgn}(\ell)\,.
\end{equation}
%


\end{document}